\documentstyle[preprint,aps]{revtex}
\begin{document}
\draft
\preprint{RU9341}
\title{Nuclear Fragmentation and Its Parallels}
\author{K. Chase and A. Mekjian}

\address{Department of Physics, Rutgers University \\
Piscataway, New Jersey 08854}
\date{\today}

\maketitle

\begin{abstract}
A model for the fragmentation of a nucleus is developed.  Parallels of
the description of this process with other areas are shown which
include Feynman's theory of the $\lambda$ transition in liquid Helium,
Bose condensation, and Markov process models used in stochastic
networks and polymer physics.  These parallels are used to generalize
and further develop a previous exactly solvable model of nuclear
fragmentation.  An analysis of some experimental data is given.
\end{abstract}

\pacs{ }

\narrowtext

\section{Introduction}

The behavior of the distribution of fragments in an intermediate
energy nuclear collision has attracted a great deal of interest.  A
previous set of papers~\cite{Mekjian1,Mekjian2,Mekjian3} proposed a
statistical model for describing this behavior.  Each possible
fragmentation outcome is given a particular probability, and the
resulting partition functions and ensemble averages are known to be
exactly solvable.  In this paper we extend that canonical ensemble
model to allow more freedom in choosing the weight associated with a
particular nuclear partition.  Formulas for the ensemble averages and
the partition function derived in earlier papers are generalized, and
a recursive formula for the evaluation of the coefficients of the
partition function is developed which is useful in computing low and
high temperature behavior of the models.

The balance of the paper is concerned with the application of these
results to models of nuclear fragmentation, as well as a comparison of
these results both to other models in physics, and to other models of
fragmentation.  An explicit parallel between this model and Feynman's
approach~\cite{Feynman} to the $\lambda$ transition in liquid Helium
is proposed.  The weight given to each possible cluster distribution
in a canonical ensemble model is shown to be similar to that used by
Feynman in the cycle class decomposition of the symmetric group.
Moreover, the main parameter, called the tuning parameter $x$
in ref.~\cite{Mekjian1}, which contains the physical
quantities associated with cluster formation, is shown to have a
correspondence with a variable in Feynman's approach related to the
cost function of moving a Helium atom from one location to another.
The variable associated with this cost function is related to
that part of the parameter $x$ which has to do with internal
excitations in a cluster, i.e., its internal partition function.
Another model which has a similar structure as the canonical ensemble
model is found in the physics of polymer formation.  Polymer
formation, when modeled by Markov processes, is a special case of the
canonical ensemble models.  Indeed, canonical ensemble models can be
derived from a particular type of Markov process models, a fact which
can be exploited when interpreting the underlying physics of
canonical ensemble models.

Other models of fragmentation are easily compared to the canonical
ensemble model.  Models based on partitioning alone
\cite{Sobotka,Aichelin} are discussed briefly.  They are
also a result of assuming a certain weight is associated with each
nuclear fragmentation outcome.  However, the choice of weight is
simpler than the models proposed here.  Models based on percolation
studies \cite{Canupi,Desbois} also have some similar features.
Markov process models \cite{Kelly} as already noted are not only
similar to canonical ensemble models, but rather are the exact same
models, simply derived from a
different, phenomenological, point of view.  This is fortuitous, as
an analogy with Markov process models can provide the basis for choosing
particular canonical ensemble models for the study of nuclear fragmentation.
A final model of fragmentation to compare the canonical ensemble model
with is the generalized canonical ensemble models, of which canonical
models are a special case.  Canonical models have many advantages over
this proposed generalization, but such generalized models
may be useful in studying exotic fragmentation situations.

A section of this paper is devoted to the discussion of the
thermodynamic functions of canonical ensemble models.  Since the
canonical models are derived from a statistical mechanics assumption, it
is appropriate to consider the computation of the typical
thermodynamic functions.  After deriving the appropriate formulas,
they are applied to the case of an ideal Bose gas in $d$ dimensions.
The critical point, present for $d>2$, is discovered by plotting the
specific heat vs. the temperature.  The connection between the zeros
of the partition function and the critical behavior is also
considered, as the recursion relations allow for the computation of
the zeros for some non-trivial cases.  The zeros for $d=2$ and $d=4$
are computed, and empirically appear to lie on simple arc-like curves.

The extended canonical models are finally applied to the problem of
nuclear fragmentation.  After reviewing the general behavior of a
number of models, which are found to vary widely in their
fragmentation behavior, we focus on a small number of models which
seem to be appropriate for an ensemble description of the
fragmentation of $^{197}_{\ 79}$Au at 0.99 GeV/amu.
Several models give excellent
results, but the statistics of the experimental data are not
sufficient to distinguish a particular model from the others considered.
One particular model, although a poor model of nuclear fragmentation,
has very interesting properties and is analyzed further.  In this
model, the fragmentation distribution favors fragments of a particular
size, with a Gaussian falloff in the distribution for larger and
smaller fragments.

The paper is organized as follows:
Section~\ref{sec:Fragmentation-models} develops the canonical model
for fragmentation and discusses parallels of this model with the
$\lambda$ transition in liquid Helium, Bose condensation
and polymer physics.
Section~\ref{sec:other-models} discusses a variety of other models for
the fragmentation of a nucleus.  Alternative partitioning models and
percolation models are reviewed briefly.  Markov process models are
introduced, and are shown to give the same distribution as the
canonical models of section~\ref{sec:Exactly-solvable-models}
when the detailed balance condition is satisfied.  Additionally,
generalized canonical models are introduced, which includes
the canonical models of section~\ref{sec:Exactly-solvable-models} as a
special case.  The thermodynamic properties of fragmentation models are
considered briefly in section~\ref{sec:Thermodynamics}.  Various
thermodynamic functions are derived, and as an application the computation
of the specific heat of a finite Bose gas is given.
The zeros of the partition
function, which can indicate the presence of phase transitions, are
also investigated for a few models.
Section~\ref{sec:General-behavior-Experimental-comparison}
discusses the behavior of the ensemble averages for various models.  A
comparison with some experimental data is made.  Concluding remarks
are in section~\ref{sec:Conclusion} and some additional calculations
are included in an appendix.

\sloppy
\section{Models of Fragmentation and Partitioning Phenomena}
\label{sec:Fragmentation-models}
\fussy

In this section, we review an approach to fragmentation and
partitioning phenomena based on the canonical ensemble of
statistical mechanics.  In this approach the fragmentation
of a nucleus, or, in general, an object, is viewed in a statistical
way with a weight given to each member in the ensemble of all possible
distributions.  Mean quantities, correlations, and fluctuations are
obtained by averaging various expressions over the ensemble using this
weight.  The model considered is not limited to descriptions of
nuclear fragmentation, and the rest of this section is devoted to
introducing other areas in physics which have used a similar type of
description.  Specifically, Feynman's description of the $\lambda$
transition in liquid Helium and an example in polymer physics are
discussed.

\subsection{Exactly Solvable Canonical Models}
\label{sec:Exactly-solvable-models}

Exactly solvable canonical models, which can be used for the
study of fragmentation and partitioning phenomena, were developed in
a previous set of papers~\cite{Mekjian1,Mekjian2,Mekjian3}.
Each partition or fragmentation is given the weight
\begin{equation}
P_{A}(\vec{n}, \vec{x}) = {A! \over Q_{A}(\vec{x})}
  \prod_{k=1}^{A} {1 \over n_{k}!}
  \left( {x_{k} \over k} \right)^{n_{k}}
\label{eq:generic-probability}
\end{equation}
where \(\vec{n} = (n_{k})_{A} = (n_{1},\ldots, n_{A})\)
is the partition vector for the fragmentation or partitioning
of the $A$ objects into $n_{k}$ clusters of size $k$,
and \(\vec{x} = (x_{k})_{A} = (x_{1},\ldots,x_{A})\) is the parameter
vector with $x_{k}$ characterizing the group or cluster of
[size $k$.  The partition vector must satisfy the constraint
\( \sum_{k=1}^{A} k n_{k} = A \) and we denote the set of all
partition vectors
\begin{equation}
N_{A} = \left\{ \forall \vec{n} \left| \sum_{k=1}^{A} k n_{k} = A
  \right. \right\}
\end{equation}
The parameter vector contains the underlying physical quantities such
as the temperature $T$ and the volume $V$.  The probability condition
for $P_{A}(\vec{n}, \vec{x})$
\begin{equation}
\sum_{\vec{n} \in N_{A}} P_{A}(\vec{n}, \vec{x}) = 1
\label{eq:conservation-of-probability}
\end{equation}
determines the canonical ensemble partition function
\( Z_{A}(\vec{x}) = Q_{A}(\vec{x})/A! \) when
eq.~(\ref{eq:generic-probability}) is substituted into
eq.~(\ref{eq:conservation-of-probability}).

Previous papers dealt with two particular models.
When all the $x_{k} = x$, the $x$ model of ref.~\cite{Mekjian1},
the partition function takes on a simple form
\begin{equation}
Q_{A}(x) = x(x+1)\cdots(x+A-1) = {\Gamma(x+A) \over \Gamma(x)}
\end{equation}
For the case $x_{1} = x y, x_{k} = x, k \ne 1$, the $x y$ model of
ref.~\cite{Mekjian3},
\begin{equation}
Q_{A}(x) = \sum_{k=1}^{A} {A \choose k} {\Gamma(x+k) \over \Gamma(x)}
  \left[x(y-1)\right]^{A-k}
\end{equation}
Detailed studies \cite{Mekjian3} show that the results of the $x$ and
$x y$ models are quite similar for all cluster sizes $k>1$.

This paper considers in detail more general forms for $x_{k}$. Here,
we explicitly show how to evaluate the partition function by simple
recursive procedures. For convenience, we rewrite
eq.~(\ref{eq:generic-probability}) by making the following substitution
\begin{equation}
{x_{k}(A,V,T) \over k} = {x(A,V,T) \over \beta_{k}}
\end{equation}
so that the dependence on the physical quantities is contained within
a single parameter, $x$, and that the thermodynamic dependence and
cluster size dependence are separable.  Then the weight is given by
\begin{equation}
P_{A}(\vec{n}, x, \vec{\beta}) = {A! \over Q_{A}(x;\vec{\beta})}
  \prod_{k=1}^{A} {1 \over n_{k}!}
  \left({x \over \beta_{k}} \right)^{n_{k}}
\label{eq:weight-x-model}
\end{equation}
This is not an unreasonable constraint on the parameters, and is
easily satisfied by many models.  For example, a previous
paper~\cite{Mekjian1} developed the result
\begin{eqnarray}
x & = & {V \over v_{0}(T)} \exp
  \left\{-{a_{\nu} \over k_{B}T} - {k_{B}T \over \varepsilon_{0}}
  {T_{0} \over T+T_{0}} \right\} \nonumber \\
\beta_{k} & = & k
\label{eq:x-of-T}
\end{eqnarray}
where $T$ is the equilibrium temperature, $V$ is the freeze out volume,
and \( v_{0}(T) = h^3/(2 \pi m_{p} k_{B} T)^{3/2} \) is the quantum volume,
with $m_{p}$ the mass of a nucleon.  The $a_{\nu}$ is the coefficient
in a simplified equation for the binding energy of a cluster of size
$k$, \(E_{B} = a_{\nu}(k-1)\). The $\varepsilon_{0}$ is the level
density parameter related to the spacing of excited levels and $T_{0}$
is a cutoff temperature for internal excitations. In a Fermi gas
model, $\varepsilon_{0}$ and the Fermi energy are related by
\( \varepsilon_{0} = 4 \varepsilon_{F}/\pi^{2} \), and
since \( \varepsilon_{F} = p_{F}^2/2 m_{p} \) can be obtained from
\( 4 (4 \pi p_{F}^3 V/3 h^3) = A \), we find that
\begin{equation}
{k_{B} T \over \varepsilon_{0}} = \left({\pi \over 12}\right)^{2/3}
  {2 m_{p} k_{B} T \over \hbar^{2}} \left({V \over A} \right)^{2/3}
\label{eq:density}
\end{equation}

The evaluation of the partition function
\( Z_{A}(x;\vec{\beta}) = Q_{A}(x;\vec{\beta})/A! \) and various
ensemble averages from the weight can be derived from the generating
function for $Z_{A}(x;\vec{\beta})$
\begin{equation}
\cal{Z}(u, x, \vec{\beta}) =
  \sum_{A=0}^{\infty} Z_{A}(x, \vec{\beta}) u^{A} = \exp
  \left\{ \sum_{k=1}^{\infty} {x \over \beta_{k}} u^{k} \right\}
\end{equation}
Using this function, it was shown in
ref.~\cite{Mekjian3} that the ensemble averaged cluster
distribution $ \langle n_{k} \rangle$ is given by
\begin{equation}
\langle n_{k} \rangle = {x \over \beta_{k}}
  {Z_{A-k}(x, \vec{\beta}) \over Z_{A}(x, \vec{\beta})}
\label{eq:expectation-nk}
\end{equation}
More generally, if we define
\begin{equation}
[z]_{k} \equiv z(z-1)\cdots(z-k+1)
\end{equation}
it was shown that
\begin{equation}
\left\langle [n_{1}]_{k_{1}} \cdots [n_{A}]_{k_{A}} \right\rangle =
  \left\{ \prod_{j=1}^{A} \left({x \over \beta_{j}} \right)^{k_{j}} \right\}
  {Z_{A-\sum_{j} j k_{j}}(x;\vec{\beta}) \over Z_{A}(x;\vec{\beta})}
\label{eq:general-ensemble-average}
\end{equation}
where \( Z_{k}(x;\vec{\beta}) = 0 \) for $k<0$.

The constraint \( \sum_{k=1}^{A} k \langle n_{k} \rangle = A \) then
leads to a simple recurrence relation for $Z_{A}(x, \vec{\beta})$
\begin{equation}
Z_{A}(x, \vec{\beta}) = {x \over A}
  \sum_{k=1}^{A} Z_{A-k}(x, \vec{\beta}) {k \over \beta_{k}}
\label{eq:Z-recurrence}
\end{equation}
with \( Z_{0}(x, \vec{\beta}) = 1 \).
Then \( Z_{1}(x, \vec{\beta}) = x/\beta_{1} \),
and so on.  We can now calculate any ensemble average of
$n_{k}$ using
eqs.~(\ref{eq:general-ensemble-average}),~(\ref{eq:Z-recurrence})

{}From the last equation we see that $Z_{A}(x, \vec{\beta})$ is a
polynomial in $x$ of order $A$.  To encourage this point of view,
we will drop the dependence on $\vec{\beta}$ from the
notation for $Z_{A}$, making the dependence tacit.  Then,
the partition function can be written as
\begin{equation}
Z_{A}(x) = \sum_{k=1}^{A} Z_{A}^{(k)} x^{k}
\label{eq:Z-polynomial}
\end{equation}
where the coefficients $Z_{A}^{(k)}$ can be determined from the
recurrence relationship as follows. The first coefficient,
$Z_{A}^{(1)}$, is determined by the last term in the recurrence relation,
\( (x/A) (A Z_{0}(x)/\beta_{A}) = x / \beta_{A} \).  So
\begin{equation}
Z_{A}^{(1)} = {1 \over \beta_{A}}
\label{eq:Zk-k=1}
\end{equation}
{}From this coefficient we can determine all the others
\begin{eqnarray}
Z_{A}(x) & = & {x \over A} \sum_{j=1}^{A} {j \over \beta_{j}}
  Z_{A-j}(x) = {x \over A} \sum_{j=1}^{A} {j \over \beta_{j}}
  \sum_{k=1}^{A-j} Z_{A-j}^{(k)} x^{k} \nonumber \\
& = & {1 \over A} \sum_{j=1}^{A} {j \over \beta_{j}}
  \sum_{k=1}^{A-j} Z_{A-j}^{(k)} x^{k+1} \nonumber \\
& = & {1 \over A} \sum_{j=1}^{A} \sum_{k=2}^{A-j+1}
  {j \over \beta_{j}} Z_{A-j}^{(k-1)} x^{k}
\end{eqnarray}
Thus all the coefficients $Z_{A}^{(k)}$ can also be obtained recursively
\begin{equation}
Z_{A}^{(k)} = {1 \over A}
  \sum_{j=1}^{A-k+1} {j \over \beta_{j}} Z_{A-j}^{(k-1)}
\label{eq:Zk-recurrence}
\end{equation}
For $k$ near $A$, this expression can be used to obtain exact results
for the coefficients.  Assuming
\( \beta_{1} = 1 \), which can always be done by
redefining $x$, $\beta_{k}$ such that $x/\beta_{k}$ is unchanged,
i.e. $x\rightarrow x/\beta_{1}$, $\beta_{k} \rightarrow \beta_{k}/\beta_{1}$
, we find
\begin{eqnarray}
Z_{A}^{(A)}   & = & {1 \over {A!}} \nonumber \\
Z_{A}^{(A-1)} & = & {1 \over {(A-2)! \beta_{2}}} \nonumber \\
Z_{A}^{(A-2)} & = & {1 \over {(A-3)! \beta_{3}}} +
                    {1 \over {2 (A-4)! \beta_{2}^{2}}} \nonumber \\
Z_{A}^{(A-3)} & = & {1 \over {(A-4)! \beta_{4}}} +
                    {1 \over {(A-5)! \beta_{2} \beta_{3}}} +
                    {1 \over {6(A-6)! \beta_{2}^{3}}} \nonumber \\
Z_{A}^{(A-4)} & = & {1 \over {(A-5)! \beta_{5}}} +
                    {1 \over {(A-6)!}}
  \left( {1 \over {2 \beta_{3}^{2}}} +
         {1 \over {\beta_{2} \beta_{4}}} \right) \\
\label{eq:Zk-largek}
            &   & + {1 \over {2(A-7)! \beta_{2}^{2} \beta_{3}}}+
                    {1 \over {24(A-8)! \beta_{2}^{4}}} \nonumber \\
Z_{A}^{(A-5)} & = & {1 \over {(A-6)! \beta_{6}}} +
                    {1 \over {(A-7)!}}
  \left( {1 \over {\beta_{5} \beta_{2}}} +
         {1 \over {\beta_{4} \beta_{3}}} \right) \nonumber \\
            &   & + {1 \over {(A-8)!}}
  \left( {1 \over {2 \beta_{4} \beta_{2}^{2}}} +
         {1 \over {2 \beta_{3}^{2} \beta_{2}}} \right) \nonumber \\
            &   & + {1 \over {6 (A-9)! \beta_{3} \beta_{2}^{3}}} +
                    {1 \over {120 (A-10)! \beta_{2}^{5}}} \nonumber
\end{eqnarray}
In general we see $Z_{A}^{(A-k)}$
depends on $\beta_{1}, \ldots, \beta_{k+1}$.
\begin{equation}
Z_{A}^{(A-k)} = \sum_{s=1}^{k} {1 \over (A-k-s)!}
  {\sum_{\begin{array}{c} \vec{n} \\ {\displaystyle \sum_{r=1}^{s}}
  j_{r} n_{j_{r}}=k+s \end{array}}}
  {1 \over \prod_{r=1}^{s} n_{j_{r}}! \beta_{j_{r}}^{n_{j_{r}}}}
\end{equation}

The recurrence relation given by eq.~(\ref{eq:Z-recurrence})
is simply solved for the case \( \beta_{k} = k \)
(as previously noted) which gives
\begin{equation}
Z_{A}(x; \beta_{k} = k) =
{1 \over A!} \sum_{k} \left| S_{A}^{(k)} \right| x^{k}
\end{equation}
where $S_{A}^{(k)}$ are Stirling numbers of the first kind.
This model was analyzed extensively in
refs.~\cite{Mekjian1,Mekjian2,Mekjian3}.

Another case which reduces to a simple polynomial is
\( \beta_{k} = 1 \) which gives
\begin{equation}
Z_{A}(x; \beta_{k} = 1) = {1 \over A} x L_{A-1}^{1}(-x)
\end{equation}
with $L_{A}^{1}(x)$ a Laguerre polynomial.  The \( \beta_{k} = 1 \)
model is considered in detail in~\cite{Gross1} as a model for
fragmentation and in~\cite{Kelly} as an example of a Markov process
model for clusterization of one dimensional objects.

A final example whose coefficients are common mathematical functions
is the choice \( \beta_{k} = k! \),
\begin{equation}
Z_{A}(x; \beta_{k} = k!) = {1 \over A!} \sum_{k=1}^{A}
  {\cal S}_{A}^{(k)} x^{k}
\end{equation}
with ${\cal S}_{A}^{(k)}$ Stirling numbers of the second kind.
This choice for the case $x=1$ was considered in detail in
ref.~\cite{DeAngelis1}.  It will also be analyzed more generally in
section~\ref{sec:General-behavior}.

For any choice of $\beta_{k}$, the recursion relation given in
eq.~(\ref{eq:Z-recurrence}) holds.  However, for some $\beta_{k}$ there
are simpler recursion relations.  For example, if $Q_{A}(x)$ is given
by an orthogonal polynomial, (e.g. \( \beta_{k} = 1 \) ), then
\( Q_{A+1}(x) = (a_{A}+ b_{A} x) Q_{A}(x) - c_{A} Q_{A-1}(x) \),
as given in Abramowitz and Stegun~\cite{Abramowitz}.
Table~\ref{tab:recursion-relations} lists some of these models.
Note that the last choice for $\beta_{k}$ in
table~\ref{tab:recursion-relations} can be related to the Catalan
numbers
\begin{equation}
C_{k} = {1 \over k+1} {2k \choose k}
\end{equation}
Specifically, \( \beta_{k} = 2^{k-1}/C_{k-1} \).  If we use Stirling's
approximation for the factorials in $\beta_{k}$ for this choice of
$\beta_{k}$, then $\beta_{k} \approx 2^{k} k^{3/2}$ for large $k$.

All the cases considered so far are special cases of some general forms
The cases \( \beta_{k} = 1, k, k/2^{k-1} {2(k-1) \choose {k-1}}^{-1} \)
can be realized from
\begin{equation}
\beta_{k} = {k! [c]^{k-1} \over [a]^{k-1} [d]^{k-1}}
\end{equation}
where $[z]^{k}$ is defined as
\begin{equation}
[z]^{k} \equiv z(z+1)\cdots(z+k-1)
\end{equation}
The case \( \beta_{k} = k! \) is a special case of
\begin{equation}
\beta_{k} = {k! [c]^{k-1} \over [a]^{k-1}}
\end{equation}

One case which does not reduce to a commonly known polynomial
and is of general interest is \( \beta_{k} = k^{\tau} \).
For example, an ideal Bose gas in $d$ dimensions can be modeled by
this choice, with \( \tau = 1+d/2 \) (see section~\ref{sec:Specific-Heat}).
In the large $A$ limit, the $Z_{A}^{(k)}$ coefficients for small
$k$ are equal to $z_{A}^{(k)}/\beta_{A}$, where the $z_{A}^{(k)}$
are only weakly dependent on $A$.  Table~\ref{tab:zk-coefficients}
gives the large $A$ limit for these coefficients.  Notice that
\(\lim_{A \rightarrow \infty} z_{A}^{(2)} = \zeta(\tau) \).

Partition functions which satisfy recurrence relationships are common
in physics.  For example, consider the ideal Boltzmann gas
\begin{equation}
Z_{A}(x) = {1 \over A!} x^{A} = {x \over A} Z_{A-1}(x)
\end{equation}
This is equivalent to the $x$ model with the following choice of parameters
\begin{eqnarray}
x & = & {V \over v_{0}(T)} \nonumber \\
\beta_{k} & = & \delta_{k1}
\end{eqnarray}
Since $\beta_{k} = 0$ for $k \ne 1$, this model has only fragments of
size one, i.e., there is no clusterization.  For a noninteracting
Boltzmann gas, this is clearly the correct behavior.

Another example of a recurrence relation in statistical mechanics is
the interacting Boltzmann gas. For this example, Feynman~\cite{Feynman}
showed that when the three body and higher order terms are neglected,
the spatial part of the recurrence relation
\begin{equation}
Z_{A+1} = V \left( 1-{a \over V} \right)^{A} Z_{A}
\end{equation}
where
\(a = \int_{0}^{\infty} \left( 1-e^{-V(r)/k_{B} T} \right)
4 \pi r^{2} dr\), with $V(r)$ the two body potential.

We now discuss some applications to other physical systems and
illustrate the parallel with fragmentation phenomenon.

\subsection{Parallel with Feynman's Approach to the
$\lambda$ Transition in Liquid Helium}
\label{sec:Feynman}

In this section we give some of the results of Feynman's
approach~\cite{Feynman} for the $\lambda$ transition which are relevant for
the analogy to be discussed.  Further details of the results quoted
can be found in~\cite{Feynman} and the references therein.

The starting point is the partition function obtained by a path
integral, given by eq.~(11.52) in ref.~\cite{Feynman}
\begin{eqnarray}
e^{-F/k_{B}T} = &&{1 \over N!}
  \left({2 \pi m' k_{B} T \over h^{2}} \right)^{3N/2}\times \nonumber \\
&&\sum_{P\in S_{N}}\int d^{3}\vec{R}_{1}\cdots d^{3}\vec{R}_{N}
  \rho(\vec{R}_{1},\ldots,\vec{R}_{N}) \nonumber \\
&&\times \exp \left\{{-{m' k_{B} T \over 2\hbar^{2}}
  \sum_{i} \left(\vec{R}_{i}-P(\vec{R}_{i})\right)^{2}} \right \}
  \label{eq:Bose-gas-Q1}
\end{eqnarray}
where $N$ is the total number of Helium atoms,
$m'$ is the effective mass,
$\vec{R_{i}}$ the coordinate of the i'th Helium atom
$\rho(\vec{R}_{1},\ldots,\vec{R}_{N})$ is the potential contribution
and $P$ is the permutation operator.
A given permutation among the particles is illustrated in
figure~\ref{fig:permutations}, and can be visualized as the atoms
being connected by a set of edges, the edges forming polygons
(cycles) of various sizes (cycle lengths).  After some
algebra and approximations, the partition function is reduced to
\begin{eqnarray}
e^{-F/k_{B} T} = \sum_{\vec{n} \in N_{N}} &\prod&_{k=1}^{N}
  {1 \over n_{k}!} \left( {N h_{k} \over k} \right)^{n_{k}} \nonumber \\
&\times& \exp \left\{ -{m'd^{2}k_{B} T \over 2 \hbar^{2}}
  \sum_{k=2}^{N} k n_{k}\right\}
\label{eq:Bose-gas-Q2}
\end{eqnarray}
where \( h_{k} = (c/k^{3/2}+1/N) l^{k} \), with $l$ the number of
nearest neighbors per lattice site for a particular choice of spatial
discretization.  This result is arrived at by a random walk argument.
Specifically, starting at a given atom, there
are $l^{k}$ random walks in $k$-steps, and the fraction of these that
are close (end up at the origin) is inversely proportional to the
volume in which the random walk is likely to end.  This volume is in
turn proportional to $k^{3/2}$.  The random walk result is then
corrected for the very large polygons that encompass a large fraction
of the sites.  This determines the form for $h_{k}$, as given above.

Comparing the results of this section with those of
section~\ref{sec:Exactly-solvable-models}, we see a strong parallel
between the model of fragmentation and the model for the $\lambda$
transition.  First, the exponent of eq.~(\ref{eq:Bose-gas-Q2}) is
identical to that internal excitation function of
eqs.~(\ref{eq:x-of-T}),~(\ref{eq:density}), up to a numerical
constant,if we neglect the cutoff temperature factor
(\(T_{0} \rightarrow \infty\))
and make substitutions for the density (\( A/V \rightarrow d^{-3} \))
and the quantum volume (\( v_{0}(T) \rightarrow d^3 \)).
Secondly $h_{k}^{-1}$ is analogous to the
parameter $\beta_{k}$ in eq.~(\ref{eq:weight-x-model}).  In fact, the
modification Feynman makes to $h_{k}$ can be well motivated in the
case of nuclear fragmentation, and we will consider the case
\( 1/\beta_{k} = a/k^{\tau}+(1-a)/k \) in
section~\ref{sec:Experimental-comparison}. Third, the partition
function has a formal structure identical to the $x y$ model.

{}From the above remarks we note two important issues in the choice of
weight given to each partition, fragmentation or grouping.  One issue
is the choice of $\beta_{k}$ and the second is the relation of $x$ to
the physical quantities.

\subsection{Polymerization Processes}
\label{sec:Polymerization}

In organic chemistry, one can model the formation of
polymers using the same models considered here, if one is
not overly concerned with the details of the molecules formed, only
in their size.  We assume that the molecules combine to form polymers
by forming bonds between molecules, up to a maximum of $f$ bonds on a
single molecule.

If existing bonds between molecules break at a rate $\kappa$ and
new bonds form at rate proportional to the number of sites available
for new bonds, then one can show (see ref.~\cite{Kelly}, chapter 8, and
references therein) that the equilibrium polymer distribution is given
by applying the $x$ model with the following parameters
\begin{eqnarray}
x & = & \kappa \nonumber \\
\beta_{k} & = & {k!((f-2)k+2)! \over ((f-1)k)!}
\label{eq:polymer-model}
\end{eqnarray}
In this case $A$ is the number of molecules, $\langle n_{k} \rangle$
is the expected
number of polymers containing $k$ molecules.  Kelly obtained this
distribution for polymer sizes by developing a Markov process model
for polymerization.  We will show later in this paper
(Section~\ref{sec:Markov-process}) that such models are equivalent to
the $x$ model with an appropriate choice for $\beta_{k}$.

The Stirling limit of $\beta_{k}$ for $f=3$ is
\begin{equation}
\beta_{k} \approx \sqrt{\pi} 2^{-2k} k^{1/2} (k+1)(k+2) \approx
  \sqrt{\pi} 2^{-2k} k^{5/2}
\end{equation}
Thus this model of polymerization is asymptotically similar to the
Bose gas choice for $\beta_{k}$ for $d=3$ for large $k$ (the term $2^{-2k}$
does not affect the distribution and can be ignored).  Since the
replacement $(k+1)(k+2) \approx k^2$ is not very good for small $k$, this
approximation is poor for large $x$, since then the configurations are
dominated by small clusters, whose behavior is determined by
$\beta_{k}$ for small $k$.  Taking \( \beta_{k} = k^{1/2}(k+1)(k+2) \) and
\( x = \kappa/\sqrt{\pi} \), figure~\ref{fig:polymer-figure} shows a
comparison of the mean number of polymers with $k$ units with the
original $\beta_{k}$ and $x$ of eq.~(\ref{eq:polymer-model}) for
$f=3, x=40$.

\section{Comparison with Other Models of Fragmentation}
\label{sec:other-models}

In this section, we discuss a number of other models of nuclear
fragmentation.  Models based on partitioning
alone~\cite{Sobotka,Aichelin}
are similar to the model outlined in
section~\ref{sec:Exactly-solvable-models} but with a simpler choice
for the  partition weight.  Percolation models~\cite{Canupi,Desbois}
are derived from far different assumptions.  Markov process models are
identical to the $x$ model, but derived from a phenomenological point
of view.  They are useful for considering what forms of $\beta_{k}$
would be appropriate for modeling fragmentation phenomenon.  Lastly,
a generalized canonical model is introduced.  Although it does not
allow for easy computation of the various ensemble averages, it may be
useful in investigating the behavior of exotic partition functions.

\subsection{Models Based on Partitioning Alone}
\label{sec:Partitioning}

Sobotka~and~Moretto~\cite{Sobotka} and
Aichelin~and~Hufner~\cite{Aichelin} discussed a model of fragmentation
based on partitioning alone with no microstate counting factor or
tuning  parameter.  In particular, their assumption is that every
partition is equally likely and that alone determines the
fragmentation.  The number of partitions of $A$ is $P(A)$ which
can be obtained from the generating function
\begin{equation}
\sum_{A=0}^{\infty} P(A) x^{A} = \prod_{A=1}^{\infty} {1 \over 1-x^{A}}
\end{equation}
This is asymptotically given by the Hardy-Ramanjuan result
\begin{equation}
P(A) \approx {1 \over 4 A \sqrt{3}} e^{\pi\sqrt{4A/3}}
\end{equation}
The number of partitions of $A$ with fixed multiplicity
\(m = \sum_{k=1}^{A} n_{k}\) is $P(A,m)$ and is given by the recurrence
relation
\begin{equation}
P(A,m) = P(A-1, m-1) + P(A-m,m)
\end{equation}
In this approach the frequency of clusters of size $k$ is
\( \langle n_{k} \rangle = P(A-k) + P(A-2k) + \ldots \)
which can be reduced to
\begin{equation}
\langle n_{k} \rangle \approx {1 \over \exp\{({\pi^2 \over 6A})^{1/2} k\} - 1}
\label{eq:percolation-nk}
\end{equation}

The above simple model of fragmentation can be generalized to include
a tuning parameter $x$ which contains the underlying physical
quantities such as volume, temperature, binding and excitation energy
associated with a fragmentation process.  In this generalization
of the models given in ref.~\cite{Sobotka,Aichelin}, the
weight given to any partition is
\begin{equation}
{x^{m} \over Q_{A}(x)}
\end{equation}
Here the $Q_{A}(x)$ is the normalization factor for this model,
and is given by
\begin{equation}
Q_{A}(x) = \sum_{m=1}^{A} P(A,m) x^{m}
\end{equation}
Once $Q_{A}(x)$ is obtained, various mean quantities can be found.
For example
\begin{equation}
\langle n_{k} \rangle = {1 \over Q_{A}(x)} \sum_{r=1}^{A} x^{r} Q_{A-r k}(x)
\end{equation}
and for $k \ne j$
\begin{equation}
\langle n_{j} n_{k} \rangle =
  {1 \over Q_{A}(x)} \sum_{r s} x^{r+s} Q_{A-r k-s j}(x)
\end{equation}
while for $k = j$
\begin{equation}
\langle n_{k} (n_{k}-1) \rangle =
  {1 \over Q_{A}(x)} \sum_{r s} x^{r+s} Q_{A-(r+s)k}(x)
\end{equation}
For large $A$ and $x A \gg 1$, $\langle n_{k} \rangle$ can be shown to approach
\begin{equation}
\langle n_{k} \rangle \approx {1 \over {1 \over x} \exp
  \left\{\left({\pi^2 x \over 6A}\right)^{1/2} k\right\} - 1}
\end{equation}
At $x = 1$, the above formula reduces to the result of
eq.~(\ref{eq:percolation-nk}), as expected.

\subsection{Percolation Models}
\label{sec:Percolation}

The $x$ model has one variable that describes the degree of
fragmentation.  The $x$ ranges from $0$ to $\infty$ as the temperature
changes over the same range.  Another approach to cluster
distributions is based on percolation which also uses one parameter.
The application of percolation to nuclear fragmentation was developed
by several groups~\cite{Canupi,Desbois}.
The percolation models are of two types: bond and
site.  The bond type assign a certain probability $p$ of having a bond
between lattice sites, while the site type assign a certain probability
of having a site occupied.  The $x$ of eq.~(\ref{eq:x-of-T})
and $x/A$ have terms which deal with volume or density effects,
and binding effects.  Also
included in $x$ are thermal effects through $v_{0}$ and internal
excitation energy considerations not present in percolation studies.

In percolation studies, the number of clusters of size $k$ is given by
\begin{equation}
\langle n_{k}(p) \rangle = {1 \over k^{\tau}} f((p-p_{c}) k^{\sigma})
\end{equation}
where $p_{c}$ is the critical probability above which an infinite
cluster exists, $f$ is a scaling function, and $\tau$, $\sigma$ are
critical exponents.

For the choice \( \beta_{k} = k \), the $x$ model has a cluster
distribution approximately given by (see ref.~\cite{Mekjian3})
\begin{equation}
\langle n_{k} (k \ll A, \epsilon \ll 1) \rangle \approx
  {x \over k} e^{-\epsilon k / A}
\end{equation}
with $x = 1+\epsilon$, giving $\tau = 1$, $\sigma = 1$.
At $\epsilon = 0$ or $x = 1$, $n_{k} = 1/k$, a hyperbolic power law
behavior.  Other choices for $\beta_{k}$ would
give a different $k$ dependence for $n_{k}$.

\subsection{Markov Process Models}
\label{sec:Markov-process}

An alternative point of view for modeling nuclear fragmentation
comes from Markov processes, which allows the underlying physical
phenomena to be reflected in the equilibrium distribution.
The idea is to consider a method by which a cluster configuration
can change into another configuration, and then to derive what the
equilibrium distribution is for such a method applied to the set of
states.  Rather than assuming what the probability for each state is,
it is derived from the distribution achieved by applying the Markov
process repeatedly.

For example, we can consider that the underlying physical processes
are the joining of two fragments to form a new larger fragment and
the splitting of a larger fragment into two smaller fragments.  Of
course, fragments joining and breaking into more than two groups are
possible,  but we will ignore that for now, assuming that those
processes are of lesser importance and will not materially affect the
overall equilibrium distribution.  We denote these processes by a
transition operator $T$ which acts as follows on $\vec{n}$, the
configuration vector.
\begin{eqnarray}
T_{l}^{j k}\vec{n} & = & (n_{1}, n_{2}, \ldots, n_{j}-1, \ldots, n_{k}-1,
\ldots, n_{l}+1, \ldots ) \nonumber \\
T_{l}^{j j}\vec{n} & = & (n_{1}, n_{2}, \ldots, n_{j}-2, \ldots,
n_{l}+1, \ldots ) \nonumber \\
T_{j k}^{l}\vec{n} & = & (n_{1}, n_{2}, \ldots, n_{j}+1, \ldots, n_{k}+1,
\ldots, n_{l}-1, \ldots ) \nonumber \\
T_{j j}^{l}\vec{n} & = & (n_{1}, n_{2}, \ldots, n_{j}+2, \ldots,
n_{l}-1, \ldots )
\end{eqnarray}
Now obviously these operators do not conserve particle number unless
$j+k = l$, so we need only consider a smaller set of operators
\begin{eqnarray}
T^{j k} & = & T_{j+k}^{j k} \nonumber \\
T_{j k} & = & T_{j k}^{j+k}
\end{eqnarray}

Suppose these processes occur at some rate. denoted by
$q(\vec{n}, \vec{n}')$ for $\vec{n}$ transforms to $\vec{n}'$.
For example,
\begin{eqnarray}
q(\vec{n}, T^{j k}\vec{n}) & = & \lambda_{j k} n_{j} (n_{k}-\delta_{j k})
\nonumber \\
q(\vec{n}, T_{j k}\vec{n}) & = & \mu_{j k} n_{j+k}
\end{eqnarray}
This is a very reasonable choice, as the probability is proportional
to the number of fragments available for such moves.  If there are
none, then the transition probability is zero, as needed.

We expect that this process when applied repeatedly to a configuration
will lead to an equilibrium configuration, at which point the rate at
which transitions occur to a new state, weighed to reflect the
equilibrium distribution of the original state, is equal to the rate at
which transitions occur back to the original state, weighed to reflect
the equilibrium distribution of the new state.
In other words, if $P_{A}(\vec{n})$ is the equilibrium
distribution, it must satisfy the detailed balance condition:
\begin{equation}
P_{A}(\vec{n}) q(\vec{n}, T^{j k} \vec{n}) =
P_{A}(T^{j k} \vec{n}) q(T^{j k} \vec{n}, \vec{n})
\end{equation}

If there exist positive numbers $c_{1}, \ldots, c_{A}$ such that
\begin{equation}
c_{j} c_{k} \lambda_{j k} = c_{j+k} \mu_{j k}
\label{eq:Markov-recurrence}
\end{equation}
then it is easy to show that the equilibrium distribution is given by
\begin{equation}
P_{A}(\vec{n}) = {1 \over Z_{A}} \prod_{k} {c_{k}^{n_{k}} \over n_{k}!}
\end{equation}
If for some choice of $\lambda_{j k}, \mu_{j k}$, we get
\( c_{k} = x/\beta_{k} \), then the models considered in
section~\ref{sec:Exactly-solvable-models} are reproduced. Once we know
$c_{k}$ we can use
eqs.~(\ref{eq:general-ensemble-average}),~(\ref{eq:Z-recurrence})
to solve for the various ensemble averages. In fact, the recursion
relationship stated above is another way of expressing the recursion
relationship relating partition functions developed in
section~\ref{sec:Exactly-solvable-models}.  Of course we still have
not produced a set of $\lambda_{j k}$, $\mu_{j k}$ which satisfy
eq.~(\ref{eq:Markov-recurrence}).  A very general solution (though not
unique) is given by the following choice
\begin{eqnarray}
\lambda_{j k} & = & \alpha f(j) f(k)\nonumber \\
\mu_{j k} & = & \beta f(j+k)
\end{eqnarray}
where $f(k)$ is any nonnegative function of $k$.
It can be easily shown that the solution to this model
(in the language of the $x$ model) is
\begin{eqnarray}
x & = & \beta/\alpha \nonumber \\
\beta_{k} & = & f(k) \label{eq:Markov-solution}
\end{eqnarray}

Thus we can always generate the $x$ model from a Markov process, using
the above prescription for $\lambda_{j k}, \mu_{j k}$.
The benefit of this approach is that the connection between the
underlying physical processes and the corresponding probability
distribution is easier to consider.
The parameters $\lambda_{j k}$ determine the association aspects of the
model, while the $\mu_{j k}$ determine the dissociation aspects.
Table~\ref{tab:Markov-models} lists a number of examples
and their solution in terms of the $x$ model.

Suppose we think of the fragmentation process as having all the
nucleons involved constrained to move in a small volume
of space for a period of time long enough for the fragments to achieve
an equilibrium.  The probability of joining two fragments should be
determined mostly by the density of the nucleons in this volume, and
the cross section of each fragment.  We expect larger
nuclei to accrete smaller nuclei due simply to their larger cross
section.  Therefore $\lambda_{j k}$ should increase monotonically
in $j$ and $k$.  All the nuclei created can break up into smaller nuclei.
We expect larger nuclei to be more unstable than smaller nuclei.
This is not strictly correct, as larger nuclei are energetically
favorable in their ground state configurations.  However, in the
aftermath of a high energy collision, the added excitation energy and
angular momentum should make larger structures unstable.
Therefore $\mu_{j k}$ should increase monotonically in $j$ and $k$ as well.
This suggests that the above example might be a reasonable model of
nuclear fragmentation, provided $f(k)$ increases monotonically.

Since all the transition operators can be generated from a smaller set of
operators, an obvious simplification of this model is to consider
the distributions generated by a restricted set of transition operators.
For example, we can restrict the Markov transition operators down to
two operators, one which adds a single nucleon to a fragment, the
other which removes a single nucleon from a fragment.
We denote these restricted operators by
\begin{eqnarray}
T^{k} & \equiv & T_{1, k-1} \nonumber \\
T_{k} & \equiv & T^{1, k-1}
\end{eqnarray}
and the restricted transition coefficients
\begin{eqnarray}
\lambda_{k} & = & \lambda_{1,k} \nonumber \\
\mu_{k}  & = &  \mu_{1,k-1}
\end{eqnarray}
Using these restricted operators any general operator (and therefore any
state) can be generated by composing the restricted operators together
in an appropriate combination.  Specifically,
\begin{eqnarray}
T^{j k}  & = &
\prod_{l=2}^{j+k} T_{j+k+2-l}
\prod_{m=2}^{j} T^{m} \prod_{n=2}^{k} T^{n} \nonumber \\
T_{j k}  & = &
\prod_{m=2}^{j} T^{j+2-m} \prod_{n=2}^{k} T^{k+2-n}
\prod_{l=2}^{j+k} T_{l}
\end{eqnarray}
Therefore if the recursion relationship given by
eq.~(\ref{eq:Markov-recurrence}) holds, it can be built up from
\begin{equation}
c_{k} \mu_{k} = c_{k-1} c_{1} \lambda_{k-1}
\label{eq:Markov-restricted-recurrence}
\end{equation}
For these models, one needs only specify the parameters $\lambda_{k}$
and $\mu_{k}$.  The full $\lambda_{j k}, \mu_{j k}$ can then generated
from the set of $c_{k}$ by applying the full recurrence relation
eq.~(\ref{eq:Markov-recurrence}).  A number of models of this type are
listed in table~\ref{tab:Markov-process2}.

Although these models give the same ensemble averages as the
more general models, they have different underlying physics.
The parameters $\mu_{k}$, $\lambda_{k}$ still determine the
dissociation and association aspects of the model; however it is
easier to discover functions that satisfy the restricted recurrence
relation given by eq.~(\ref{eq:Markov-restricted-recurrence})
than it is to satisfy the full recurrence relation
of eq.~(\ref{eq:Markov-recurrence}).
In other words, these models allow more freedom to adjust the
dissociation and association rates.  For instance the model
\begin{eqnarray}
\lambda_{k} & = & \alpha k^\sigma \nonumber \\
\mu_{k} & = & \beta k^\tau
\end{eqnarray}
which has the solution
\begin{eqnarray}
x & = & {\beta \over \alpha} \nonumber \\
\beta_{k} & = & {k!^{\tau} \over (k-1)!^{\sigma}}
\end{eqnarray}
is not easily produced from an obvious choice of
parameters $\lambda_{jk}$, $\mu_{jk}$.

The above model conforms to the idea introduced earlier of the nucleons
combining and dissociating as a gas in a limited volume.  Single
nucleons can detach themselves from larger nuclei, and larger nuclei
can accrete single nucleons.  If we assume that the nucleons
automatically combine if they get close enough, then the association rates
should be roughly proportional to the surface area of the
large fragment, i.e \( \sigma = 2/3 \).  If we assume that nucleons
near the surface of the large fragments leave at a constant rate, then
the dissociation rates should also be a surface phenomenon, and \( \tau
= 2/3 \), which implies \( \beta_{k} = k^{2/3} \).  We will see in
section~\ref{sec:Experimental-comparison} that this model does not
reproduce experimental data well.  Instead models with $\sigma =
\tau \ge 1$ produce better results.

\subsection{Generalized Canonical Models}
\label{sec:General-canonical-models}

As a final example of a model of fragmentation, we discard the notion
that the model must be derived from any particular choice of weight.
Indeed, the usual process of choosing a weight and then deriving its
partition function can be reversed.  A partition function can be
chosen, and a weight scheme that generates such a partition function
can be computed.  It is important to note that the choice of a
partition function is not sufficient for fixing such a weight scheme.
Many different choices for a weight lead to the same partition function
and only additional assumptions can fix the weight scheme.
For example, the model given by the weight
\begin{equation}
P_{A}(\vec{n}, x) = {A! \over Q_{A}(x)} \prod_{k=1}^{A}
  {1 \over n_{k}!} \left( {x \over k} \right)^{n_{k}}
\end{equation}
and by the weight
\begin{equation}
P_{A} \left(\vec{n}, x \left| \right. m = \sum n_{k} \right) =
  {x^{m} \over Q_{A}(x)}
  {\left| S_{A}^{(m)} \right| \over \cal{S}_{A}^{(m)}}
\end{equation}
both give the same partition function
\begin{equation}
Q_{A}(x) = \sum_{m=1}^{A} \left| S_{A}^{(m)} \right| x^{m}
\end{equation}

Suppose that we are given a partition function
\begin{equation}
Q_{A}(x) = \sum_{k=1}^{A} Q_{A}^{(k)} x^{k} =
  \sum_{\{\vec{n}\}} W_{A}(\vec{n})
\end{equation}
and we want to determine a weight scheme that generates this partition
function.  For the canonical model with \( \beta_{k} = k \), the
weight is given by \( W_{A}(\vec{n}) = M_{2}(\vec{n}) x^{m} \), where
\( m = \sum n_{k} \), \( M_{2}(\vec{n}) = A!/\prod n_{k}! k^{n_{k}} \).
An obvious generalization of this weight would be the choice
\begin{equation}
W_{A}\left( \vec{n} \in N_{A} | m = \sum n_{k} \right)
  = W_{A}^{(m)} A! \prod_{k=1}^{A} {1 \over n_{k}!}
  \left( {x \over k} \right)^{n_{k}}
\end{equation}
i.e., $W_{A}(\vec{n})$ is equal to the standard canonical model weight,
up to a factor that depends only on the number of fragments.
For this weight we can easily show
\begin{eqnarray}
\sum_{ \left\{ \vec{n} \in N_{A} \left| \right. m = \sum n_{k} \right\} }
  W_{A}(\vec{n}) &=& W_{A}^{(m)} \left| S_{A}^{(m)} \right| x^{m}
  \nonumber \\
&=& Q_{A}^{(m)} x^{m}
\end{eqnarray}
which implies that
\begin{equation}
W_{A}^{(m)} = {Q_{A}^{(m)} \over \left| S_{A}^{(m)} \right|}
\end{equation}
or that the weight is given by
\begin{equation}
W_{A} \left( \vec{n} \in N_{A}  \left| \right. m = \sum n_{k} \right)
  = {Q_{A}^{(m)} \over \left| S_{A}^{(m)} \right|} A!
  \prod_{k=1}^{A} {1 \over n_{k}!} \left( {x \over k} \right)^{n_{k}}
\end{equation}

Now if we could write down the generating function for $Q_{A}(x)$
we could compute $\langle n_{k} \rangle$ as was done in
section~\ref{sec:Exactly-solvable-models}.
However, in general there is no generating function for $Q_{A}(x)$.
We can calculate \( \langle m \rangle \equiv \sum \langle n_{k}
\rangle \) exactly, however.
\begin{equation}
\langle m \rangle = {x \over Q_{A}(x)} {\partial Q_{A} \over \partial x}
\end{equation}

The $\langle n_{k} \rangle$ results are not entirely inaccessible.
They can be obtained by a Monte Carlo simulation of the
partition function.  Note first that the partition function can be
expressed as a sum over the permutation group.
\begin{eqnarray}
Q_{A}(x) & = & \sum_{
  \left\{ \vec{n} \in N_{A} \left| \right. m = \sum n_{k} \right\}}
  {A! \over \prod_{k=1}^{A} n_{k}! k^{n_{k}}}
  {Q_{A}^{(m)} \over \left| S_{A}^{(m)} \right| } x^{m} \nonumber \\
& = & \sum_{p \in S_{A}} { Q_{A}^{(m(p))} \over
  \left| S_{A}^{(m(p))} \right|} x^{m(p)} =
  \sum_{p \in S_{A}} e^{-S(p)}
\end{eqnarray}
with
\begin{eqnarray}
S(p) = \log \left| S_{A}^{(m(p))} \right| - \log Q_{A}^{(m(p))} -
  m(p) \log x
\end{eqnarray}
We can simulate this action over the set of permutation using the
Metropolis algorithm, with any of a number of choices for transition
functions.  For example, if we denote $p_{k}$ as the action of the
permutation operator on $k$, one possible transition function is
\( p_{k} \leftrightarrow p_{k+1} \) for a uniformly random choice of $k$.

The choice of $Q_{A}(x)$ cannot be completely arbitrary.  In the
large $x$ limit, it must reduce to \( Q_{A}(x) = x^{A} + O(x^{A-1}) \),
which fixes \( Q_{A}^{(A)} = 1 \).  Similarly, the small $x$ limit
fixes the choice of $Q_{A}^{(1)}$.  Other than these considerations
and the requirement that that the coefficients be positive,
there are no other restrictions on the partition function.  Indeed,
some fairly exotic choices can be made.

One example of a generalized canonical model is given by the partition
functions generated by
\begin{equation}
Q_{A+1}(x) = Q_{A}({ax + b \over cx +d}) (cx+d)^{A+1} \Theta_{A}
\end{equation}
with \( Q_{1}(x) = x\), and $\Theta_{A}$ a constant which corrects
$Q_{A}^{(A)}$.  Models of this type satisfy a simple recurrence
relation, but that recurrence relation is quite different than the
canonical recurrence relation given in eq.~(\ref{eq:Z-recurrence}).
Another model, even more exotic, is given by
\begin{equation}
Q_{2A}(x) = Q_{A}({qx(x+1) \over x+q}) (x+q)^{2A} \Theta_{A}
\end{equation}
Models such as this are interesting in studies of the roots of
partition functions~\cite{Derrida,Peitgen}, since the computation of large
numbers of zeros for such partition functions is easily accomplished.
The distribution of roots in the complex plane can have a fractal
character when the requirement that the coefficients be positive is
relaxed.  The case $q=-1$, shown in figure~\ref{fig:fractal-zeros},
reveals one such complex distribution of roots.
The roots lie on the boundary of a series of copies of the
Mandelbrot set.  A discussion of zeros
of the partition function for canonical models will be made in
section~\ref{sec:Zeros}.

\newcommand{\dxdt}{{\partial x \over \partial t}}
\newcommand{\dxdv}{{\partial x \over \partial v}}
\newcommand{\dsqxdtsq}{{\partial^{2} x \over \partial t^{2}}}
\newcommand{\dQdt}{{\partial Q_{A}(x) \over \partial t}}
\newcommand{\dQdv}{{\partial Q_{A}(x) \over \partial v}}
\newcommand{\dQdx}{{\partial Q_{A} \over \partial x}}
\newcommand{\dsqQdxsq}{{\partial^{2} Q_{A} \over \partial x^{2}}}
\newcommand{\dsqQdtsq}{{\partial^{2} Q_{A}(x) \over \partial t^{2}}}

\section{Thermodynamic Properties of Fragmentation Models}
\label{sec:Thermodynamics}

In this section we consider the various thermodynamic functions that
can be computed from the partition functions obtained from the weight
given in eq.~(\ref{eq:weight-x-model}).  As an example, the specific
heat for a finite Bose gas is computed.  Finally, a discussion of
phase transitions leads us to consider the location of the roots of
the partition function on the complex plane for several models.

\subsection{Thermodynamic Functions}
\label{sec:Thermo-functions}

In section~\ref{sec:Exactly-solvable-models} we introduced a model
with the thermodynamic variable confined to a single parameter $x$.
Making this assumption allows us now to simply calculate the
thermodynamic functions of such partition functions.  Since
\begin{equation}
e^{-F(A, V, T)/k_{B} T} = Q_{A}(x) = A! \sum_{\{n_{k}\}}
\prod_{k=1}^{A} {1 \over n_{k}!} \left( {x \over \beta_{k}} \right)^{n_{k}}
\end{equation}
is the partition function for a thermodynamic system with
\( x = x(A,V,T) \), it is straightforward to calculate the
thermodynamic functions from the free energy.
First, let us introduce dimensionless variables for $T$ and $V$
\begin{equation}
t \equiv {T \over T_{1}},  v \equiv {V \over V_{1}}
\end{equation}
$T_{1}$ and $V_{1}$ are arbitrary reference points, but we will find
it convenient (in nuclear fragmentation) to use the values
\begin{equation}
k_{B} T_{1} = a_{\nu},  V_{1} = {4 \over 3}\pi r_{0}^{3} A
\end{equation}
where $a_{\nu}$ defines the scale of binding energies,
and $r_{0}$ is the classical radius of a nucleon.

We can express the various thermodynamic functions in terms of
$\langle m \rangle$, $\langle m^2 \rangle$, $x$ and its derivatives.
The calculations are simple, and here we quote the results for the
dimensionless entropy, pressure, free energy, energy, and specific heat.

\begin{eqnarray}
s & = &  S/k_{B} = {-1 \over k_{B}}
        \left({\partial F \over \partial T}\right)_{V} \nonumber \\
  & = &  \ln Q_{A}(x) + \langle m \rangle \left({t \over x} \dxdt \right) \\
p & = & P V_{1}/k_{B} T_{1} = {V_{1} \over k_{B} T_{1}}
        \left({-\partial F \over \partial V}\right)_{T} \nonumber \\
  & = & \langle m \rangle \left( {t \over x} \dxdv \right) \\
f & = & F/k_{B}T_{1} = {1 \over k_{B} T_{1}}
        \left(-k_{B} T \ln Q_{A}(x) \right) \nonumber \\
  & = & -t \ln Q_{A}(x) \\
u & = & U/k_{B}T_{1} = {1 \over k_{B} T_{1}}
        \left(F + T S\right) = f + t s \nonumber \\
  & = & \langle m \rangle \left( {t^{2} \over x} \dxdt \right) \\
c_{V} & = & C_{V}/k_{B} = {-T \over k_{B}} \left(
  {\partial^{2} F \over \partial T^{2}} \right)_{V} \nonumber \\
  & = & \langle m \rangle \left(2 {t \over x} \dxdt -
  \left( {t \over x} \dxdt \right)^{2}
  +{t^{2} \over x} \dsqxdtsq \right) \nonumber \\
  &  & + \left(\langle m^{2} \rangle-\langle m \rangle^{2} \right)
          \left({t \over x} \dxdt \right)^{2}
\label{eq:specific-heat}
\end{eqnarray}
where the ensemble averages of $m$, $m^2$ are obtained as follows
\begin{eqnarray}
\langle m \rangle & = &  \left\langle \sum_{k} n_{k} \right\rangle =
  {x \over Q_{A}(x)} \dQdx \\
\langle m(m-1) \rangle & = &  \left\langle \sum_{j} n_{j}
  \left( \sum_{k} n_{k}-1 \right) \right\rangle \nonumber \\
& = & {x^2 \over Q_{A}(x)} \dsqQdxsq
\end{eqnarray}

\subsection{Specific Heat of a Bose Gas}
\label{sec:Specific-Heat}

We can apply the above expressions to obtain the specific heat
of a finite Bose gas.  A Bose gas in $d$ dimensions can be modeled
by the $x$ model, with
\begin{eqnarray}
x & = & {V \over v_{0}(T)} \nonumber \\
\beta_{k} & = & k^{1+d/2} \label{eq:Bose-gas-model}
\end{eqnarray}
where \( v_{0}(T) = (h^2/2 \pi m_{p} k_{B} T)^{d/2} \).
Combining eqs.~(\ref{eq:Bose-gas-model}),~(\ref{eq:specific-heat}) we
arrive at a simple formula for the specific heat
\begin{equation}
c_{V} = {d \over 2} \langle m \rangle +
  {d^2 \over 4} (\langle m^2 \rangle - \langle m \rangle^{2})
\end{equation}
For $d>2$, there is a phase transition (Bose-Einstein condensation)
in the infinite particle limit, which can be seen  as a cusp in the
specific heat at the critical point, $x_{c} = A/\zeta(d/2)$.  For finite
gases, the partition function is smooth, so there is no cusp.
However the specific heat does reach a maximum near the critical point,
suggesting the cusp will onset in the the large particle limit.  This
behavior is illustrated in figure~\ref{fig:specific-heat-Bose-gas}.

This model could be taken as a model of nuclear fragmentation,
with a different expression for $x$.  Indeed, it will be shown in
section~\ref{sec:Experimental-comparison} that models with
\( \beta_{k} = k^{\tau} \) with \( \tau \ge 1 \) are fairly good models of
fragmentation.  The analog of Bose-Einstein condensation into the
ground state is the formation of the largest cluster, that is the
``condensation'' of the nucleons into the single cluster with $k=A$,
which is called the fused mode in ref.~\cite{Mekjian1}.

The canonical ensemble partition function is obtained from
eq.~(\ref{eq:Z-recurrence}).  The polynomial associated with this
partition function, given by eq.~(\ref{eq:Z-polynomial}), has the
following coefficients which are obtained by applying
eqs.~(\ref{eq:Zk-k=1}),~(\ref{eq:Zk-recurrence}), and
(\ref{eq:Zk-largek}).
\begin{eqnarray}
Z_{A}^{(A)}   & = & {1 \over A!} \nonumber \\
Z_{A}^{(A-1)} & = & {1 \over 2^{1+d/2}(A-2)!} \nonumber \\
Z_{A}^{(A-2)} & = & {1 \over 3^{1+d/2}(A-3)!} +
              {1 \over 2(2^{1+d/2})^{2}(A-4)!} \\
Z_{A}^{(A-3)} & = & {1 \over 4^{1+d/2}(A-4)!} +
              {1 \over 2^{1+d/2}3^{1+d/2}(A-5)!} \nonumber \\
              & + & {1 \over 6(2^{1+d/2})^{3}(A-6)!} \nonumber \\
     & \vdots & \nonumber \\
Z_{A}^{(2)} & \approx & {\zeta(1+d/2) \over A^{1+d/2}} \nonumber \\
Z_{A}^{(1)}   & = & {1 \over A^{1+d/2}} \nonumber
\end{eqnarray}
These coefficients can be used to obtain the behavior of various
thermodynamic functions in the low and high temperature limits.

In the following subsection, we will consider another way of
obtaining the phase transitions of a partition function.  Using the
results obtained in earlier sections, we attempt to calculate the
zeros of the partition function for various $A$.

\subsection{Zeros of the Partition Function and Phase Transitions}
\label{sec:Zeros}

The canonical partition function $Z_{A}(x)$ is a polynomial of order
$A$ in $x$ with positive coefficients.  For example, for
\( \beta_{k} = k \), \( Z_{A}(x) = x(x+1)\cdots(x+A-1)/A! \)
which when expanded gives
\begin{equation}
Z_{A}(x) = {1 \over A!}\sum_{m=1}^{A} \left| S_{A}^{(m)} \right| x^m
\end{equation}
with $S_{A}^{(m)}$ Stirling numbers of the first kind.
By a theorem of Gauss, a
polynomial of order A has A roots or zeroes in the complex plane.
For positive coefficients, no real roots are on the positive
real-axis, which is also the physical meaningful axis.  The above
example has its roots at the $x=0$ and the negative integers
\( x=-1, -2, \ldots, -A+1 \).

Complex roots correspond to an extension of the real temperature into
the complex plane.  Lee and Yang, in their discussion of phase
transitions, showed that such transitions manifest themselves as
zeros of the partition function approach the real positive axis
as the thermodynamic limit $A \rightarrow \infty$ is approached.
Taking the logarithm of the partition function to obtain the free
energy can then lead to a singularity.

To illustrate the above remarks, we consider the example of the 2-d
Ising model on an $m \times n$ lattice.  Kauffman~\cite{Kauffman}
showed that the partition function for this model is given by
\begin{equation}
Z_{mn} = \prod_{r=1}^{m} \prod_{s=1}^{n}
\left\{ \left( {1+v^2 \over 1-v^2} \right)^2
- {2 v f_{r s} \over 1-v^2} \right\}
\end{equation}
where
\begin{eqnarray}
f_{r s} & = & \cos(2 \pi r/m) + \cos(2 \pi s/n) \nonumber \\
v  & = & \tanh (J/k_{B}T)
\end{eqnarray}
In this case the zeroes of the partition function are located on
two circles in the complex $v$-plane, namely
\(v = \pm 1 + \sqrt{2} e^{i \theta} \).  The physically meaningful
domain of the $v$-plane is the part of the real line $v \in (0,1)$
( assuming $J > 0$).
The zeros of the partition function approach this domain at one
point, $v = (-1 + \sqrt{2})$.  This implies there should be a
phase transition when
\begin{equation}
k_{B} T_{c} = {J \over \tanh^{-1} (-1 + \sqrt{2})} =
   {2 J \over \log (1+\sqrt{2})}
\end{equation}
which is the commonly known value for the critical temperature.

Now consider the zeros of the $x$ model partition function
in the complex $x$-plane.  The physically meaningful domain of $x$,
the part with positive temperature, is the positive real axis.  So for a
phase transition to manifest itself in the infinite particle limit,
the zeros of the partition function must approach the positive real axis.
The $\beta_{k} = k$ model therefore has no phase transition,
for in the infinite particle limit, the roots of the partition
function are zero and the negative integers, which never approach the
real temperature domain.  Another example we consider is
\( \beta_{k} = k^{\tau} \) for the cases \( \tau = 2, 3 \).
This corresponds to an ideal Bose gas in two and four dimensions by
eq.~(\ref{eq:Bose-gas-model}).  So we expect that the zeros should
approach the physically meaningful domain for large $A$ for the case
$\tau=3$, but not for the case $\tau=2$.
Figures~\ref{fig:complex-roots-2d}~and~\ref{fig:complex-roots-4d}
illustrate the zeros for the models \(\beta_{k} = k^2\) and
\(\beta_{k} = k^{3}\) for $A=25,50,75,100$.  These graphs suggest that
for both models, the roots lie on simple curves.  Whether these curves
close on the positive real axis is not clear from these small $A$
results.  The roots near the negative real axis scale with $A$,
which suggests that the crossing point on the positive real axis should
scale with $A$ as well.  This agrees with the known behavior of the
critical point, which is given by \( x_{c}/A = 1/\zeta(d/2) \).  The fact
that the curve will not close for $d=2$ is not evident from the
graph, however.

\sloppy
\section{General Behavior of Models and Comparison with Experimental Data}
\label{sec:General-behavior-Experimental-comparison}
\fussy

In this section we consider the general behavior of $\langle n_{k}
\rangle$ for various  $\beta_{k}$ and how well these models proposed
actually fit some experimental data obtained from heavy ion collider
experiments~\cite{Waddington}.  This data was obtained from emulsion
experiments for $^{197}_{\ 79}$Au at 0.99 GeV/amu.  415 events were
recorded and identified by the charges of the fragments, i.e.
each event is represented by \( \vec{n} = (n_{1},\ldots,n_{79}) \)
where $n_{z}$ represents the number of fragments with charge $z$ in a
given event.  Ensemble averages were obtained by averaging over all
events.

Since the experimental method only measured the electric charge of
fragments leaving the collision, there is some question about how
applicable are models developed considering only nucleons with no
separation into protons and neutrons.  Since the
nuclear force treats protons and neutrons nearly identically, and the
models proposed derived from a combinatorial viewpoint, the models
should be identical whether one includes the neutrons or not.  In
fact, it can be shown that for the simple \( \beta_{k} = k \) model,
that the results are the same whether one considers $Z$ nucleons coming
out, or whether one considers $A$ nucleons coming out, but only the
$Z$ protons can be followed, such that one must sum over the possible
neutron configurations to obtain the expectation values.

\subsection{General Behavior of the Models}
\label{sec:General-behavior}

Because all the models must satisfy \( \sum_{k} k n_{k} = A \), there are
restrictions on the form of the distribution.  If we graph
\(\langle m_{k} \rangle = k \langle n_{k} \rangle \)  vs. $k$, then
the area under the curve must be equal to $A$. For different choices
of $x$ and $\beta_{k}$, the area will be distributed differently.  In
this subsection we discuss typical distributions for these models.

All $x$ models have simple behaviors which are easy to obtain
at large and small $x$.  For small $x$, all models will produce
$\langle m_{k} \rangle$ with most of the area under $\langle m_{A}
\rangle$.  This is because for small $x$, the partition function is
given almost entirely by the term proportional to $x$.  This implies
that one fragment is the most likely outcome.  For $k \ne A$,
$\langle n_{k} \rangle$ and $\langle m_{k} \rangle$ are proportional
to $x$ since
\begin{equation}
\langle n_{k} \rangle = {x \over \beta_{k}} {Z_{A-k}(x) \over Z_{A}}
  \approx x {\beta_{A} \over \beta_{k} \beta_{A-k}}
\label{eq:nk-small-x}
\end{equation}
For large $x$, all models will produce $\langle m_{k} \rangle$ with
most of the area under $\langle m_{1} \rangle$.  This is because for
large $x$, the partition function is given almost entirely by the term
proportional to $x^{A}$.  So $A$ fragments is the most likely outcome.
For $k \ne 1$, $\langle n_{k} \rangle$ and $\langle m_{k} \rangle$ are
proportional to $x^{1-k}$ since
\begin{equation}
\langle n_{k} \rangle = {x \over \beta_{k}} {Z_{A-k}(x) \over Z_{A}}
  \approx  {x^{1-k} \over \beta_{k}} {A! \over (A-k)!}
\label{eq:nk-large-x}
\end{equation}

The model \( \beta_{k} = k \) was discussed in an earlier set of papers.
For small $x$, most of the area is near $k=A$, as expected.
For $x<1$, $\langle m_{k} \rangle$ is monotonically increasing.
At $x=1$, $\langle m_{k} \rangle = 1$ for all $k$.
For $x>1$, $\langle m_{k} \rangle$ is monotonically decreasing.
At large $x$, most of the area is near $k=1$, as expected.
This is shown in figure~\ref{fig:mk-tau1.0}.

For models with \( \beta_{k} = 1 \), for $x \ll 1$ most of the
area is below $k=A$.  As $x$ increases, some area is distributed along
the rest of the graph, mostly around $k=A/2$.  As $x$ keeps
increasing, the area continues to be redistributed, until most of it
is distributed about a point $k<A/2$.  At large $x$ it takes on
the usual distribution.  This is illustrated in figure~\ref{fig:mk-tau0.0}.

For the model \( \beta_{k} = k^{\tau} \), with $0 < \tau < 1$,
the behavior is very similar to \( \beta_{k} = 1 \). For small $x$,
$\langle m_{k} \rangle$ increases monotonically,with most of the
area near $k=A$.  As $x$ increases, the right hand side is
diminished till the graph attains two turning points,
a local minimum near $k=A$, and a local maximum at
a point $k<A/2$.  The local minimum soon disappears, and the
local maximum migrates left till it reaches $k=1$ at large $x$.
This behavior is illustrated in figure~\ref{fig:mk-tau0.5}.

The models \( \beta_{k} = k^{\tau} \), $\tau>1$ are all very similar.
For small $x$, $\langle m_{k} \rangle$ starts monotonically decreasing,
reaches a minimum, then near $k=A$ rises rapidly.
As $x$ is increased, the small $k$ behavior is unchanged,
but eventually the large $k$ part turns downward.
Thus for a small range of $x$, the graphs
have two turning points. As $x$ continues to increase,
the local maximum eventually disappears and the typical
large $x$ behavior onsets.  This behavior is shown in
figure~\ref{fig:mk-tau2.0}.

The model \( \beta_{k} = k! \), for $x \ll 1$, $\langle m_{k} \rangle$
is mostly under $k=A$, as expected.  From eq.~(\ref{eq:nk-small-x}), we see
$\langle n_{k} \rangle \approx x {A \choose k}$ for $k \ne A$, so the
remainder of the area is binomially distributed about $k=A/2$. As $x$
increases, the amount of area under $k=A$ diminishes, the balance
appearing around $k \approx A/2$ in a binomial or Gaussian distribution.
Once most of the area has disappeared from $k=A$, the Gaussian
distribution at the center moves to the left as $x$ is increased.
For very large $x$, most of the area is under $k=1$, as expected.

A simplified description of the \( \beta_{k} = k! \) model can be
obtained by making the following approximations.  For a given $x$, the
partition function is strongly peaked about a particular number of
fragments $m$  (i.e. $Q_{A}(x) \approx Q_{A}^{(m)} x^{m}$).  To estimate
this value of $m$ we first use the approximation for the Stirling
numbers of the 2nd kind, $S_{A}^{(m)} \approx m^{A}/m!$, in the expression
for the partition function.  This approximation is very good for
$m \ll A$, and is only off by a factor of two for $m \approx A/2$.  Using
this approximation, we can calculate which term in the partition
function dominates, i.e. for what value of $m_{0}$, $Q_{A}^{(m_{0})} x^{m_{0}}$
is maximal.  We discover the following nonlinear equation for $m_{0}$
\begin{equation}
m_{0} = x e^{A/m_{0}} = m_{0}(A, x)
\end{equation}
In obtaining this result, Stirling's approximation for
\(m! \approx m^{m+1/2} e^{-m} \sqrt{2 \pi}\) was used. The distribution
$\langle n_{k} \rangle$ is obtained from eq.~(\ref{eq:expectation-nk}).
If we make the approximation
\begin{equation}
\langle n_{k} \rangle = {x \over k!} {Z_{A-k}(x) \over Z_{A}(x)} \approx
  {A \choose k} m_{k}^{-k} f_{A}(x)
\end{equation}
where \(m_{k} = m_{0}(A-k,x)\).  We now make the assumption that
\begin{equation}
m_{k} \approx m_{0}(A-\langle k \rangle,x) \equiv \bar{m}
\end{equation}
The value of $f_{A}(x)$ can be obtained by imposing
\( \sum k \langle n_{k} \rangle = A \), which gives
\begin{equation}
\langle n_{k} \rangle \approx {A \choose k} p^{k} (1-p)^{A-k} (1+\bar{m})
\label{eq:approximate-model}
\end{equation}
where \( p = 1/(\bar{m}+1) \).  This distribution is binomial, with
\begin{equation}
\langle k \rangle = {A \over \bar{m}+1}
\end{equation}
Using this result we can determine the equation for $\bar{m}$.
\begin{equation}
\bar{m} = m_{0}(A - {A \over \bar{m}+1}, x) =
  m_{0}(A{\bar{m} \over \bar{m}+1}, x) = x e^{A/(\bar{m}+1)}
\end{equation}
Figure~\ref{fig:factorial-approximation}
compares the exact behavior with this approximation.  From the figure
we see that $\langle n_{k} \rangle$ is reasonably well
described by this approximation.

\subsection{Experimental Comparisons}
\label{sec:Experimental-comparison}

Nuclear fragmentation has a characteristic distribution which is met
generically by only a few of the above models.  For the experiment
we will analyze, $\langle m_{k} \rangle$ drops, then rises.  This
suggests models with $\beta_{k} = k^{\tau}$ with $\tau>1$ might be
satisfactory if one $x$ is used.  Models with two or more $x$'s are
considered in ref.~\cite{Mekjian3}.  The large $k$ behavior is
somewhat indeterminate.  It could be rising or falling; there are not
enough events to determine the behavior accurately.

Fits were made to $\log \langle n_{k} \rangle$, dropping from the
experimental distribution any $\langle n_{k} \rangle$ that were zero
due to insufficient statistics.  A previous paper~\cite{Mekjian3}
showed that for $x=0.3$, the \( \beta_{k}=k \) model gives a fairly
good fit.  A better fit is obtained by using two or more $x$'s.
Here, we consider the models proposed in the previous subsection,
as well as a mixed model analogous to Feynman's choice
for the $\lambda$ transition~\cite{Feynman},
\( 1/\beta_{k} = a/k + (1-a)/k^{\tau} \), with various $\tau$.
These results are shown in table~\ref{tab:experimental-fits}, and in
figure~\ref{fig:best-fits}

\section{Conclusions and Summary}  \label{sec:conclusion}
\label{sec:Conclusion}

This paper presents a detailed investigation of a set of exactly
solvable canonical ensemble models of fragmentation processes and
discusses some of its parallels with other areas.  Specifically,
parallels between the description of the fragmentation process and
other areas are developed which include Feynman's approach to the
$\lambda$ transition in liquid helium, Bose condensation, and Markov
process models used in stochastic networks and polymer physics.

The partition functions derived from various weights given to each
member of the canonical ensemble, are shown to be polynomials in a
parameter $x$.  Simple recurrence procedures are developed for
obtaining the partition function and the coefficients in the
associated polynomials.  The variable $x$, called a tuning parameter,
contains the underlying physical quantities associated with the
description of the the different processes considered.  For example,
for the ideal Boltzmann gas, ideal Fermi-Dirac gas, and ideal Bose
gas, $x$ involves the thermodynamic variables $V$ (volume) and $T$
(temperature) through the quantum volume $v_{0}(T)$.
For fragmentation, $x$ also includes
binding energy and excitation energy coefficients associated with
cluster formation and in the Feynman description of the $\lambda$
transition, $x$ contains the cost function of moving a helium atom
from one position to another.  This cost function for the $\lambda$
transition is shown to be related to that part of $x$ in fragmentation
processes that involves internal excitations in a cluster.  The length
of the cycle of a permutation in the symmetrization of the Bose system
wavefunction is the analog of the cluster size $k$ and the cycle class
decomposition of the symmetric group is equivalent to the partitioning
or grouping of the original $A$ objects into clusters of various sizes.

Besides the tuning parameter $x$, the weight given to each member of
the ensemble contains a quantity called $\beta_{k}$ which gives the
cluster size or cycle length $k$ dependence of this weight.
Various choices for $\beta_{k}$ are considered, and a wide range of
different types of behavior can be found for different choices for
$\beta_{k}$.  The Bose gas in $d$-dimensions has $\beta_{k} =
k^{1+d/2}$ and Bose condensation exists for $d > 2$.  The Feynman
approach to the $\lambda$ transition is based on $\beta_{k}^{-1} =
ak^{-5/2} + (1-a)k^{-1}$.  A previous model of fragmentation
\cite{Mekjian1,Mekjian2,Mekjian3} used $\beta_{k} = k$,
a choice leading to very simple results due to some theorems in
combinatorial analysis.  The partition function for this last choice
is a simple polynomial in $x$ whose coefficients are the signless
Stirling numbers of the 1st kind.  A more general form, $\beta_{k} =
k^{\tau}$, is also investigated here.  For a partition function that
leads to a phase transition, $\tau > 2$.  Also considered here is the
choice $\beta_{k} = k!$, which is shown to have some interesting
properties.  The partition function `in this case involves Stirling
numbers of the second kind, and the distribution of fragments obtained
from this partition function represents systems which split into large
equal or nearly equal size pieces.  Specifically, this choice gives
rise to Brownian or binomial type distributions of clusters whose peak
is centered around cluster sizes $A/m$ where $A$ is the number of
objects and $m$ is the mean multiplicity, which is a function of $x$.

A consideration of the thermodynamics of fragmentation systems led
to an investigation of the behavior of the partition function when $x$
is a complex number.  In particular, the zeros of the partition function
are studied in the complex plane and the connection with the Lee-Yang
theorems and phase transitions are investigated for various choices of
$\beta_{k}$.  More complex iterative models of the partition function
are proposed whose distribution of zeros are fractal sets.

This paper also presents an alternative point of view for modeling
fragmentation processes which is based on Markov process models.
Markov process models give a picture of the underlying physical
processes that lead to the cluster formation and break up.  The
relationship of this approach to that based on the canonical ensemble
is discussed.

Finally, some experimental data is investigated.  Various choices for
the quantity $\beta_{k}$ are considered in our analysis.  The
statistics of the data are not sufficient to distinguish the various
possible $\beta_{k}$'s considered.

\acknowledgments
This work supported in part by the National Science Foundation
Grant \# NSFPHY92-12016. One author (KC) wishes to thank Rutgers
University Excellence Graduate Fellowship for providing support during
part of this research.

\appendix

\section*{Generating Functions and Recurrence Relations}
\label{sec:Generating-function}

In this section we consider a general procedure for generating the
canonical partition function $Q_{A}(x, \vec{g})$ from a generating
function given by
\begin{equation}
\cal{Q}(u, x, \vec{g}) = \exp \left\{ x \sum_{k=1}^{\infty}
  g_{k} u^{k} \right\} = \sum_{A=0}^{\infty}
  Q_{A}(x,\vec{g}) {u^{A} \over A!}
\end{equation}
Here the $g_{k}$'s are arbitrary functions of $k$.  Letting
$y = x \sum g_{k} u^{k}$, expanding $\exp y = 1 + y + y^{2}/2 +
\ldots$ and collecting all terms with equal powers of $u$ gives
\begin{equation}
Q_{A}(x,\vec{g}) = \sum_{\Pi_{A}(\vec{n})} M_{3}(\vec{n}) x^{m}
  \prod_{k=1}^{A} \left( g_{k} k! \right)^{n_{k}}
\end{equation}
where $M_{3}(\vec{n}) = A!/\prod n_{k}! {k!}^{n_{k}}$ and consequently
\begin{equation}
Q_{A}(x, \vec{g}) = \sum_{\Pi_{A}(\vec{n})}
  {A! x^{m} \over \prod n_{k}! (g_{k}^{-1})^{n_{k}}}
\end{equation}
Thus $\beta_{k} = g_{k}^{-1}$.  For example,

\begin{eqnarray}
Q_{1}(x, \vec{g}) & = & x g_{1} \nonumber \\
Q_{2}(x, \vec{g}) & = & 2! g_{2} x + g_{1}^{2} x^{2} \nonumber \\
Q_{3}(x, \vec{g}) & = & 3! g_{3} x + 3 g_{1} (2! g_{2}) x^{2} +
  g_{1}^{3} x^{3} \\
Q_{4}(x, \vec{g}) & = & (4! g_{4}) x + \left[
  3(2! g_{2})^{2} + 4g_{1}(3! g_{3}) \right] x^{2} \nonumber \\
  &&+6 g_{1}^{2} (2! g_{2}) x^{3} + g_{1}^{4} x^{4} \nonumber
\end{eqnarray}

Using a procedure in Riordan~\cite{Riordan}, the following recurrence
relationship is obtained
\begin{equation}
Q_{A+1}(x, \vec{g}) = \left\{ xg_{1} + \sum_{s=1}^{A} (s+1) g_{s+1}
  {d \over dg_{s}} \right\}Q_{A}(x, \vec{g})
\end{equation}

\begin{figure}
\caption{Permutations among particles.  Left hand side of graph shows
a group of particles and a permutation operator as it would act on the
particles.  Right hand side gives cluster interpretation of the same
permutation.}
\label{fig:permutations}
\end{figure}

\begin{figure}
\caption{The behavior of $\langle m_{k} \rangle = k \langle n_{k}
\rangle$ for the polymer model with A=50.
(1) is the exact model given by $\beta_{k} = k!(k+2)! / (2k)!$,
(2) is the approximate model given by the Stirling limit
$\beta_{k} = k^{1/2} (k+1)(k+2)$.}
\label{fig:polymer-figure}
\end{figure}

\begin{figure}
\caption{Zeros of the partition function $Z_{A}(x)$ with $q=-1$ }
\label{fig:fractal-zeros}
\end{figure}

\begin{figure}
\caption{Specific heat of a finite Bose gas for d=3.  $k_{B} T_{1} = 8$
MeV in this figure.}
\label{fig:specific-heat-Bose-gas}
\end{figure}

\begin{figure}
\caption{Zeros of the partition function $Z_{A}(x)$ for the choice
$\beta_{k} = k^{2}$ scaled by $1/A$ (i.e. root $x$ is plotted at $x/A$).
The cases $A=25, 50, 75, 100$ are shown.}
\label{fig:complex-roots-2d}
\end{figure}

\begin{figure}
\caption{Zeros of the partition function $Z_{A}(x)$ for the choice
$\beta_{k} = k^{3}$ scaled by $1/A$ (i.e. root $x$ is plotted at $x/A$).
The cases $A=25, 50, 75, 100$ are shown.}
\label{fig:complex-roots-4d}
\end{figure}

\begin{figure}
\caption{The behavior of $\langle m_{k} \rangle$ for the choice
$\beta_{k}=k$, $A=50$ at various $x$.}
\label{fig:mk-tau1.0}
\end{figure}

\begin{figure}
\caption{The behavior of $\langle m_{k} \rangle$ for the choice
$\beta_{k}=1$, $A=50$ at various $x$.}
\label{fig:mk-tau0.0}
\end{figure}

\begin{figure}
\caption{The behavior of $\langle m_{k} \rangle$ for the choice
$\beta_{k}=k^{1/2}$, $A=50$ at various $x$.}
\label{fig:mk-tau0.5}
\end{figure}

\begin{figure}
\caption{The behavior of $\langle m_{k} \rangle$ for the choice
$\beta_{k}=k^{2}$, $A=50$ at various $x$.}
\label{fig:mk-tau2.0}
\end{figure}

\begin{figure}
\caption{The behavior of $\langle m_{k} \rangle$ for the choice
$\beta_{k} = k!$, $A=50$ at various $x$, with a comparison to the
approximate model considered in the text.}
\label{fig:factorial-approximation}
\end{figure}

\begin{figure}
\caption{$\log \langle n_{k} \rangle$ vs. $k$ for best fits. Line (1) is for
$\beta_{k}=k$.  Lines (2), (3), (4), are for
$1/\beta_{k} = a/k +(1-a)/k^{\tau}$,
with (2) $\tau = 2$, (3) $\tau = 2.5$, (4) $\tau = 3$. }
\label{fig:best-fits}
\end{figure}

\begin{table}
\caption{Recursion relations for various $\beta_{k}$}
\label{tab:recursion-relations}
\begin{tabular}{cl}
$\beta_{k}$ & Recursion relation \\
\tableline
$k$ & \( Q_{A+1} = (x+A) Q_{A} \) \\
$k!$ & \( Q_{A+1} = (x + x {d \over dx}) Q_{A} \) \\
$1$ & \( Q_{A+1} = (x + 2A) Q_{A} - A(A-1) Q_{A-1} \) \\
${k \over 2^{k-1}} {2(k-1) \choose {k-1}}^{-1}$ &
\(Q_{A+1} = (2A-1) Q_{A} + x^{2} Q_{A-1} \) \\
\end{tabular}
\end{table}

\begin{table}
\caption{Values of $z_{A}^{(k)}$ in the large $A$ limit}
\label{tab:zk-coefficients}
\begin{tabular}{cccccc}
$\tau$ & $z_{A}^{(1)}$ & $z_{A}^{(2)}$ & $z_{A}^{(3)}$ & $z_{A}^{(4)}$ &
$z_{A}^{(5)}$ \\
\tableline
3/2 & 1.000 & 2.612 & 3.412 & 2.971 & 1.941 \\
2   & 1.000 & 1.645 & 1.353 & 0.742 & 0.305 \\
5/2 & 1.000 & 1.342 & 0.899 & 0.402 & 0.135 \\
3   & 1.000 & 1.202 & 0.723 & 0.289 & 0.0870 \\
\end{tabular}
\end{table}

\begin{table}
\caption{Markov Process Models}
\label {tab:Markov-models}
\begin{tabular}{cccc}
$\lambda_{j k}$ & $\mu_{j k}$ & $x$ & $\beta_{k}$ \\
\tableline
$\alpha$ & $\beta$ & $\beta/\alpha$ & $1$ \\
$\alpha (j k)^\tau$ & $\beta (j+k)^\tau$ & $\beta/\alpha$ & $k^\tau$ \\
$\alpha$ & $\beta {j+k \choose j}^\tau$ & $\beta/\alpha$ & $k!^\tau$ \\
\end{tabular}
\end{table}

\begin{table}
\caption{Restricted Markov process models}
\label{tab:Markov-process2}
\begin{tabular}{cccc}
$\lambda_{k}$ & $\mu_{k}$ & $x$ & $\beta_{k}$ \\
\tableline
$\alpha$ & $\beta$ & $\beta/\alpha$ & $1$ \\
$\alpha k$ & $\beta k$ & $\beta/\alpha$ & $k$ \\
$\alpha$ & $\beta k$ & $\beta/\alpha$ & $k!$ \\
$\alpha k^{\sigma}$ & $\beta k^{\tau}$ & $\beta/\alpha$ &
${k!^{\tau} \over (k-1)!^{\sigma}}$ \\
\end{tabular}
\end{table}

\begin{table}
\caption{Fits of Experimental Data to various Models}
\label{tab:experimental-fits}
\begin{tabular}{ccccc}
$\beta_{k}$           & $x$     & $a$     & $\tau$ & $\sigma^{2}$ \\
\tableline
$k$                             & 0.296   & n/a    & n/a   & 78.58 \\
$k^{\tau}$                      & 1.895   & n/a    & 1.817 & 63.10 \\
${k \over a + (1-a)k^{1-\tau}}$ & 0.93745 & 0      & 1.5   & 65.61 \\
${k \over a + (1-a)k^{1-\tau}}$ & 1.96263 & 0.0587 & 2     & 62.34 \\
${k \over a + (1-a)k^{1-\tau}}$ & 2.61313 & 0.0747 & 2.5   & 61.55 \\
${k \over a + (1-a)k^{1-\tau}}$ & 3.0803  & 0.0732 & 3     & 62.20 \\
\end{tabular}
\end{table}

\end{document}